# Thermally induced structural modification in the Al/Zr multilayers

Qi Zhong,[1] Shuang Ma,[1] Zhong Zhang,[1] Runze Qi,[1] Jia Li,[1] Zhanshan Wang,[1, *] and Philippe Jonnard[2]

[1]MOE Key Laboratory of Advanced Micro-Structured Materials, Institute of Precision Optical Engineering, Department of Physics, Tongji University, Shanghai 200092, China
[2]Laboratoire de Chimie Physique – Matière Rayonnement, UPMC Univ Paris 06, CNRS UMR 7614, 11 rue Pierre et Marie Curie, F-75231 Paris cedex 05, France

**ABSTRACT:** The effect of increasing temperature on the structural stability and interactions of two kinds of Al/Zr (Al(1%wtSi)/Zr and Al(Pure)/Zr) multilayer mirrors are investigated. All Al/Zr multilayers annealed from 200 ℃ to 500 ℃, were deposited on Si wafers by using direct–current magnetron sputtering technology. A detailed and consistent picture of the thermally induced changes in the microstructure is obtained using an array of complementary measurements including grazing incidence X-ray reflectance, atomic force microscope, X–ray diffraction and high-resolution transmission electron microscopy. The first significant structural changes of two systems are observed at 250 ℃, characterized by asymmetric interlayers appears at interface. At 290 ℃, the interface consisted of amorphous Al-Zr alloy is transformed to amorphous Al-Zr alloy and cubic ZrAl$_3$ in both systems. By 298 ℃ of Al(1%wtSi)/Zr and 295 ℃ of Al(Pure)/Zr multilayers, the interfacial phases of Al-Zr alloy transform completely into polycrystalline mixtures of hcp-ZrAl$_2$ and cubic-ZrAl$_3$, which smooth the interface boundary and lower the surface roughness in the multilayers. Up to 500 ℃, the multilayer structure still exists in both systems, and the differences between the asymmetric interlayers are much larger in the multilayers. Finally, we discuss the transformation from symmetric to asymmetric in the annealing process for other systems.

## 1. INTRODUCTION

Multilayers are useful in many important technological and scientific fields, such as X-ray applications[1, 2], layer-by-layer assembly[3,4], electrical interconnects [5]. While the performance of x-ray multilayers designed as reflection mirrors under intense synchrotron radiation are greatly influenced by the power load larger than 100 W/mm. Since multilayers of different combinations have shown inherently metastable materials, the optical, structural performances and stability under intense x-ray flux and annealing processing are therefore of considerable interest. In particular, the annealing process instead of real intense radiation is often used to investigate the thermally induced structure modification in the scientific research[6-10]. For the Mo-Si systems, they has provided the most efficient reflection mirrors in the region of 13-30nm, while the thermal stability and interactions of Mo/Si multilayers have studied since they are used for soft x-ray mirrors in hostile environments. Based on the analysis of grazing incidence X-ray reflectance (GIXR), X–ray diffraction (XRD) and high-resolution electron microscopy, an increase in the width of the amorphous interlayer is started at 400 ℃. After 550 ℃, there is transition happened in the Mo layers, which the phase transforms from Mo (bcc) to Mo$_5$Si$_3$ (hcp) and MoSi$_2$ (tetragonal). The intermediate-phase formation (Mo$_5$Si$_3$ and MoSi$_2$) will shift the diffraction angles of the x-ray scattering peaks and influence the optical and structural performances of the multilayers. While the amorphous Si layer still remain intact but present a structural instability. Up to 800 ℃,

---

the tetragonal MoSi$_2$ destroys the multilayer structure completely [11, 12]. However, since the Al-based multilayers have been used for the reflective mirrors, there is few data [13-16] to present the thermal stability and interactions of the Al/Zr multilayers designed for used as reflection mirrors in the region of 17-19nm.

In our previous works [13], we present the thermal stability of one as-deposited and five annealed (100 ℃, 200 ℃, 300 ℃, 400 ℃, 500 ℃) Al(1%wtSi)/Zr multilayers. Based on the fits of the GIXR and near-normal incident extreme ultraviolet (EUV) reflection measurements, the symmetric and asymmetric interlayers in four-layer models were used to present the interfacial structure before and after 300 ℃. The X-ray emission spectroscopy (XES) and XRD measurements revealed that the Al(1%wtSi)/Zr multilayer had a stable structure and performance up to 200 ℃, which had the symmetric amorphous interlayers in the multilayers. After 300 ℃, the interfacial phases of Al-Zr alloy transformed from amorphous to polycrystalline，which degraded the optical contrast of the multilayer structure, and increased the interfacial roughness. Based on the simulation of GIXR, the multilayer structure was not fully destroyed by the polycrystalline Al-Zr compound at 500 ℃. The previous works just described the thermal stability of Al(1%wtSi)/Zr, but not accurately present what temperature of the transition from amorphous to polycrystalline occurs at interfaces. We also found the symmetry of interlayers could transform from symmetric to asymmetric with the increasing temperature for the first time, but the details explanations were not concluded in the previous works, because of lacking the corresponding compositions of Al-Zr alloy at different annealing temperatures. Based on the reference for Si-based [17-21], Mg-based [22-25] and Al-based multilayers [13-16 and 26], the transformation for the symmetry of interlayers in the annealing process was just found in the Al-Zr multilayers up to now. However, we think the transformation may not be only appropriate for the Al-Zr systems; the transformation from symmetric to asymmetric in the annealing process for other systems should be present in the following research.

Therefore, in this paper, we focus on the effect of different reaction temperatures, the formation of variable materials between Al and Zr layers in Al/Zr (Al(1%wtSi)/Zr and Al(Pure)/Zr) systems, and the transformation from symmetric to asymmetric in the annealing process for other systems. After a brief description of the experimental process (Sect. 2), three reaction temperatures for each system were characterized by GIXR, atomic force microscope (AFM) and XRD in the Sect. 3.1. To fully describe variable composition in the Al/Zr multilayers (Sect. 3.2), we simulate the XRD measurements. And a detailed and consistent picture of the thermally induced changes in the microstructure is obtained by using the transmission electron microscope (TEM) measurements. The transformation from symmetric to asymmetric in the annealing process for other systems is present in the Sect. 3.3. Finally, we conclude the different reaction points, variable compositions at interface and the transformation from symmetric to asymmetric in the annealing process for other systems (Sect. 4).

## 2. EXPERIMENT

All Al/Zr (Al(1%wtSi)/Zr and Al(Pure)/Zr) multilayer samples consisting of 40 bi-layers, are deposited on the Si wafers, by using direct–current magnetron sputtering technology [13, 27-30], under the base pressure $4.5\times10^{-5}$Pa. The sputtering gas was Ar with purity of 99.999%, and the gas pressure was held constantly at $1.20\pm0.02$mTorr (0.16Pa). The targets of zirconium (99.5%) and silicon doped aluminum (Al(1%wtSi)) or aluminum (99.999%, Al(Pure)) with diameter of 100 mm were used. The gamma value is 0.33, and the periodic thicknesses of Al(1%wtSi)/Zr and Al(Pure)/Zr multilayers are 9.56 nm and 9.35nm, respectively. To evaluate the thermal reactions of Al(1%wtSi)/Zr and Al(Pure)/Zr multilayers, the thirteen multilayer samples for each system, except for room temperature (RT) sample,

were annealed at temperatures of 200, 250, 280, 285, 290, 295, 298, 300, 305, 310, 400 and 500 ℃ in a vacuum furnace for 1 h, respectively. After annealing, the samples were cooled to room temperature naturally in the vacuum furnace under a base pressure of $3.5 \times 10^{-4}$ Pa.

For characterizing interfacial structure, the GIXR was performed by using a Cu $K_\alpha$ source (λ=0.154 nm), and the fitting data were simulated by the Bede Refs software (genetic algorithm) [31]. While the surface roughness was measured with a Veeco, MultiMode SPM scanning probe microscope, operated in AFM mode. The XRD measurements provide identification of crystalline phases present in the modified layer along with structural changes during the annealing process. To verify the interfacial reactions of two Al/Zr systems during the annealing process, the transmission electron microscope (TEM, FEI Tecnai $G^2$ F20) was used on the specimen prepared by focused ion beam (FIB) etching using in the Materials Analysis Technology Ltd.

## 3. RESULTS AND DISCUSSION

### 3.1 Reaction temperature

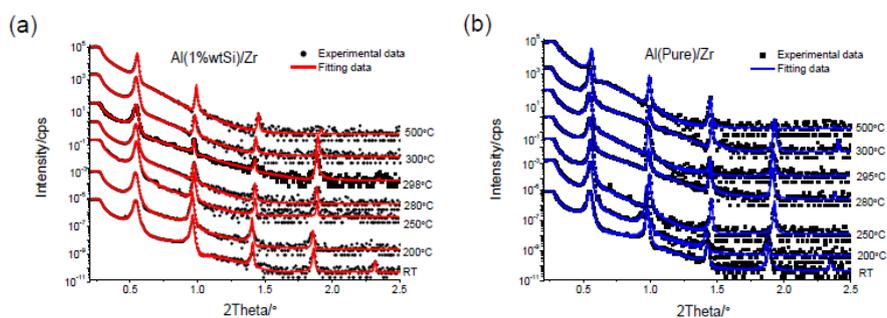

Figure 1. Comparison between the GIXR experimental (black dots) and fitted curves (color curves) with Al(1%wtSi)/Zr (a) and Al(Pure)/Zr multilayers (b) for the RT, 200, 250, 298-a/295-b, 300 and 500 ℃ samples.

As a first step, we have investigated interfacial structure, surface roughness and crystal size of the Al(1%wtSi)/Zr and Al(Pure)/Zr multilayers in different annealing temperatures, looking for the reaction temperatures in the multilayers. In the previous work [13], we use the symmetric and asymmetric interlayer models to present the interfacial structure before and after 300 ℃. Therefore, we fit the GIXR measurements of Al(1%wtSi)/Zr and Al(Pure)/Zr multilayers with the same models, respectively (Figure 1). From the fitting results in the Figure 1(a), the symmetric model can be used up to 200 ℃, which the interfacial widths are 1.5nm in both interlayers. At 250 ℃, the asymmetric model brings into use to simulate the measurements. From 250 to 298 ℃, the interfacial widths of two interlayers (Zr-on-Al and Al-on-Zr) keep at 2.0nm and 1.6nm, respectively. But the interfacial roughness of Zr-on-Al interlayer is decreased from 0.81 to 0.60nm, and that of Al-on-Zr interlayer is also changed from 0.7 to 0.5nm. From 300 ℃, the interfacial widths of two interlayers increases to 2.5nm and 2.0nm, and the roughnesses increases to 0.95nm and 0.89nm, respectively. Up to 500 ℃, the roughnesses of Zr-on-Al and Al-on-Zr interlayers are about 2.17nm and 1.10nm. While in the Al(Pure)/Zr multilayers, the interfacial width and roughness have the similar situation from 250 to 295 ℃ in Figure 1(b). From 300 ℃, both the interfacial width and roughness are increased with the annealing temperature. Up to 500 ℃, the Al and Zr layers still keep the multilayer structure.

To characterize the first reaction temperature at 250 ℃ in both multilayers, the changes in the periodic length of the annealed multilayer samples are derived from GIXR measurements (not all measurements are shown in Figure.1). In Figure 2, the periodic length of each multilayer sample is normalized by those at the RT conditions. The periodic length of the multilayer slightly decreases with the annealing temperature of Al(1%wtSi)/Zr samples (Figure 2(a)). The relative change in the periodic length of the multilayer annealed at 200 ℃ is about 0.2%. From 250 ℃, it sharply decreases to 0.5%, and up to 500 ℃ the change is about 1.8% of the initial value. In the Figure 2(b), the relative change of Al(Pure)/Zr multilayers at 200 ℃ is also about 0.2%, but from 250 ℃, the slope of the changing trend is higher than that in the Al(1%wtSi)/Zr multilayers. The change is about 0.6% at 250 ℃, and drops to 2.4% at 500 ℃. Based on the GIXR measurements, we can found that the reaction between Al and Zr layers starts at 250 ℃ , which may be influenced by large interdiffusion in both Al(1%wtSi)/Zr and Al(Pure)/Zr multilayers. It should be noted that more and more amorphous Al-Zr alloy grows into Al and Zr layers at 250 ℃, which could disturb the original balance in the interface boundary before 200 ℃. That is Zr diffuses and reacts more rapidly with Al than Al diffuses and reacts with Zr [15], which induces the asymmetric interlayers appeared in the multilayers. Thus, it is reasonable to suppose that the first reaction temperature is at 250 ℃. However, the changing trends of two systems are much different from 200 to 500 ℃, which the relative change of Al(Pure)/Zr multilayer at 500 ℃ is much higher than that of Al(1%wtSi)/Zr system. It means there is a large interdiffusion between Al and Zr layers in Al(Pure)/Zr multilayers, and the deterioration of Al(Pure)/Zr multilayer structure is much worse than that of Al(1%wtSi)/Zr.

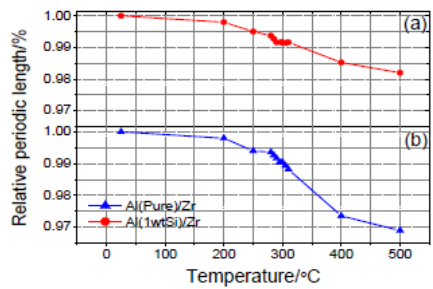

Figure 2. Relative periodic length of the Al(1%wtSi)/Zr (a, red curve) and Al(Pure)/Zr multilayers (b, blue curve) as a function of annealing temperatures from RT to 500 ℃. The periodic length was normalized by those of the samples before annealing.

Table 1. Surface roughness of the Al(1%wtSi)/Zr and Al(Pure)/Zr multilayer samples derived from the AFM measurements.

| Samples | Temperature /℃ | RT | 200 | 250 | 280 | 285 | 290 | 295 | 298 | 300 | 305 | 310 | 400 | 500 |
|---|---|---|---|---|---|---|---|---|---|---|---|---|---|---|
| Al (1%wtSi) /Zr | Before annealing(nm) | 0.86 | 0.85 | 0.85 | 0.86 | 0.85 | 0.85 | 0.86 | 0.87 | 0.86 | 0.86 | 0.86 | 0.85 | 0.86 |
| | After annealing(nm) | 0.86 | 1.05 | 1.02 | 0.99 | 0.96 | 0.95 | 0.92 | 0.72 | 0.86 | 0.84 | 0.97 | 1.04 | 1.05 |
| Al (Pure) /Zr | Before annealing(nm) | 0.80 | 0.78 | 0.83 | 0.81 | 0.83 | 0.89 | 1.18 | 1.14 | 0.78 | 0.75 | 0.71 | 0.70 | 0.75 |
| | After annealing(nm) | 0.80 | 0.93 | 0.96 | 0.92 | 0.93 | 0.99 | 1.05 | 1.06 | 0.83 | 0.86 | 0.83 | 0.82 | 0.89 |

The surface roughnesses of the Al(1%wtSi)/Zr and Al(Pure)/Zr multilayers before and after annealing are characterized with AFM (Table 1), measured over a 5 μm × 5 μm area. In order to eliminate the effect of the roughness before annealing, the relative surface roughnesses of different annealing samples are used, which are derived from Table 1 (shown in Figure 3(a)). The surface roughness of each multilayer sample is normalized by those at the as-deposited conditions. For the relative change of surface roughness in Al(1%wtSi)/Zr multilayers, it is first increased with the annealing temperature to 200 ℃. From 250 ℃, the change of surface roughness decreased with the increasing temperature. At 298 ℃, it drops drastically to 0.88% of the initial value, which shows a lowest point in the curve. After 298 ℃, the change increases with the temperature, up to 500 ℃, the change is about 1.30%. Compared the Al(1%wtSi)/Zr with Al(Pure)/Zr samples, the roughnesses in different samples are varied from 0.70nm to 1.18nm before annealing, which would influence the values after annealing. But the relative change has the similar situation with Al(1%wtSi)/Zr multilayers before 295 ℃, which also have a lowest point at 295 ℃. Starting from 295 ℃, the relative roughness increases with the annealing temperature, At 500 ℃, the change is about 1.18 %, which is lower than that in Al(1%wtSi)/Zr. This phenomenon might be influenced by the crystal size in Al layer. To sum up, the both curves of Al(1%wtSi)/Zr and Al(Pure)/Zr multilayers show different changing trends before and after the temperature at 298 ℃ and 295 ℃, respectively, which might have a reaction between Al and Zr layers at those turning points. Thus, it is reasonable to suppose that another reaction temperatures for Al(1%wtSi)/Zr and Al(Pure)/Zr multilayers are at 298 ℃ and 295 ℃, respectively.

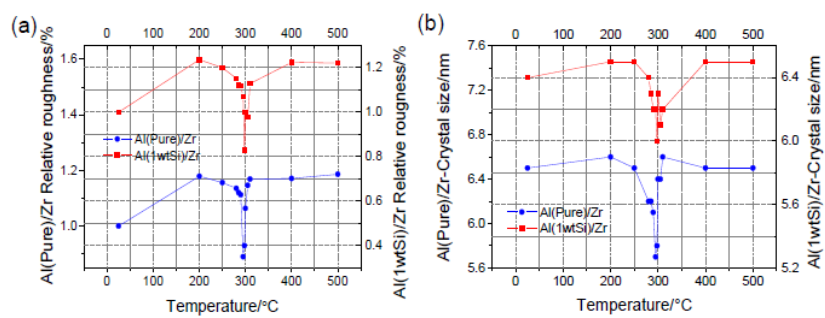

Figure 3. a: Relative surface roughness of the Al(1%wtSi)/Zr (red curve) and Al(Pure)/Zr multilayers (blue curve) as a function of annealing temperatures from RT to 500 ℃, which were derived from Table 1. The surface roughness was normalized by those of the samples before annealing. b: The crystal sizes of all annealing samples in two systems were derived from the XRD measurements.

Our previous works show that the surface and interfacial roughness were influenced by the crystallization of Al layer [27, 28]. Therefore, to identify the changes in interfacial and surface roughness, the XRD measurements were carried out in two Al/Zr multilayers. Considering the Scherrer formula [32], the crystal sizes of Al<111> in different annealing temperatures were shown in Figure 3(b). For the Al(1%wtSi)/Zr systems, the crystal size slightly increased with the annealing temperature before 200 ℃. From 250 ℃, the crystal size begins to decrease, which means the interdiffusion becomes larger in the interface, and increase the interfacial width (Figure 1(a)). At 290 ℃, the crystal size is much lower than the thickness of Al(1%wtSi) layer (6.3nm). That is to say, the FWHM is wider than that before 290 ℃. Based on the thermal reactions in Al/Zr thick multilayers (the thickness of the multilayer is beyond 100 nm) [15], the wider FWHM shows the Al<111> may not be the only phase around 38.7 °. There may have some other phases, which means there is a reaction between Al and Zr layers at interface. At 298 ℃, the crystal size is the lowest value (6.0nm), consist with the lowest point in the relative surface change in Figure 3(a). After 298 ℃, the crystal size increases with annealing temperature to 500 ℃. The crystal size is 6.6nm. For the Al(Pure)/Zr systems, the changing trend is similar with that in Al(1%wtSi)/Zr in annealing process. After 290 ℃, the crystal size is smaller than the thickness (6.2nm) of Al(Pure) layer, which has a lowest point (5.7nm) at 295 ℃. At 500 ℃, the crystal size is 6.5nm. Based on the XRD results in two Al/Zr systems, we can see that the new material could appear in the interface at 290 ℃. And the changes of the crystal size could explain the variable situation of interfacial (Figure 1) and surface roughness (Figure 3(a)), which shows the changes in crystallization of Al were correlated with the similar changes in both interfacial and surface roughness. The different crystal sizes at 500 ℃ may be influenced by the different thickness of Al layer in two systems. While because of the presence of Si, the turning point at 298 ℃ in Al(1%wtSi)/Zr is larger than that at 295 ℃ in Al(Pure)/Zr. We can see that the Si, even in a small proportion, would not only change the interfacial structure [17], but also influence the reaction temperature in the multilayers.

### 3.2 The variable structure in the annealing process

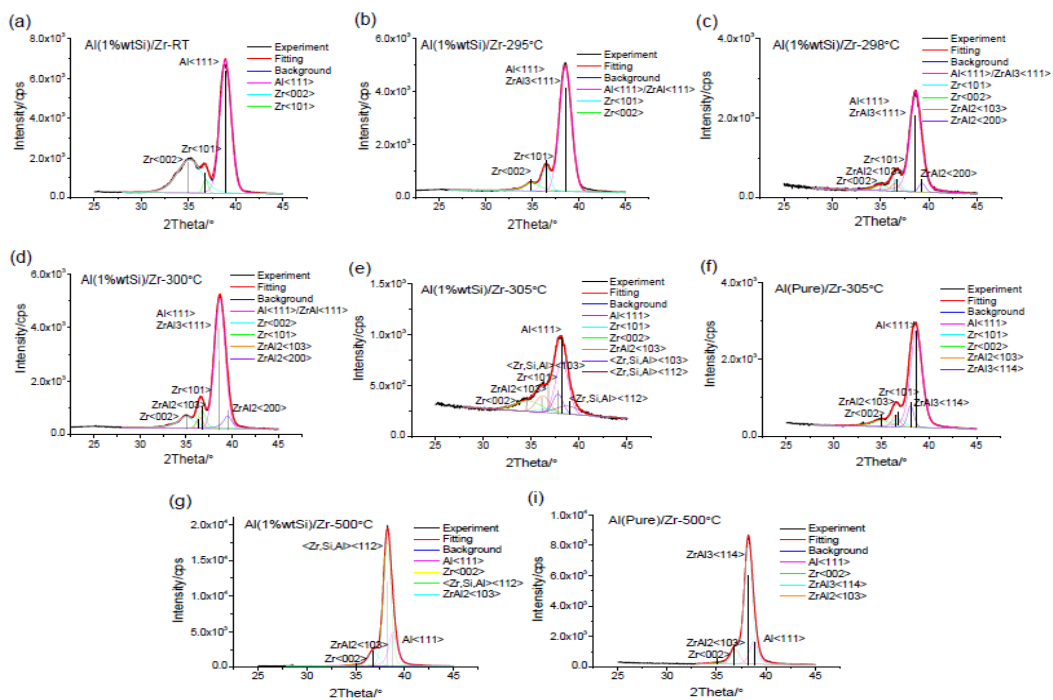

Figure 4 Fits of XRD measurements in Al(1%wtSi)/Zr (a: RT, b:295 ℃, c:298 ℃, d: 300 ℃, e: 305 ℃ and g: 500 ℃) and Al(Pure)/Zr (f: 305 ℃ and i: 500 ℃) spectra, black points: experiment results, red lines: recomposing signal, blue lines: background, color lines: different phase peaks.

In the previous section, we have found three reaction temperatures in two Al/Zr systems (250 ℃, 290 ℃ and 298 ℃ for Al(1%wtSi)/Zr, 250 ℃, 290 ℃ and 295 ℃ for Al(Pure)/Zr), which means the interfacial structure would change at the different reaction temperatures. To further confirm the reaction temperatures in both Al/Zr multilayers, the XRD measurements were fitted by Peak fit software [33, 34], which could present the detailed compositions of interfacial structure in the annealing process. We keep 2theta parameter unchanged, based on the reference values in the PDF-2004 database: 38.79 ° (cubic-Al<111>), 38.78 ° (cubic-$ZrAl_3$<111>), 39.54 ° (hcp-$ZrAl_2$<200>), 36.72 ° (hcp-$ZrAl_2$<103>), 37.94 °(tetragonal-$ZrAl_3$<114>), 38.29 °(tetragonal<Zr,Al,Si><112>), and 37.79 °(tetragonal<Zr,Al,Si><103>). While the parameters of peak height and FWHM are variable during the fitting process. Figure 4 shows the fits of XRD measurements in Al(1%wtSi)/Zr (a: RT, b: 290 ℃, c: 298 ℃, d: 300 ℃, e: 305 ℃ and g: 500 ℃) and Al(Pure)/Zr (f: 305 ℃ and i: 500 ℃) spectra. Before 300 ℃, the composition of Al(Pure)/Zr is similar with that of Al(1%wtSi)/Zr, so that we just describe the fitting data of Al(1%wtSi)/Zr in details. But after 300 ℃, the results of Al(1%wtSi)/Zr and Al(Pure)/Zr are presented separately. In the as-deposited (RT) state (Figure 4(a)), the multilayers consist of alternating layers of polycrystalline Al and Zr and thin, interfacial layers of amorphous Al-Zr alloy [13, 15 and 16]. Because the amorphous layers grow more into the Al and Zr layers, the interface consist of amorphous Al-Zr alloy has transformed to amorphous Al-Zr alloy and cubic ZrAl$_3$ at 295 ℃ (The curves of 290 ℃ is similar with that of 295 ℃). After 1h at 298 ℃ (Figure 4(c)), there are two more phases hcp ZrAl$_2$<200> and hcp ZrAl$_2$<103> appeared at interface, which influence the peaks of Zr<101> and Al<111> separately. At this point, we can get that the interface consists of amorphous Al-Zr alloy and cubic ZrAl$_3$ is transformed to hcp ZrAl$_2$ and cubic ZrAl$_3$, which also have the similar point at 295 ℃ in the Al(Pure)/Zr multilayers (the curve is not shown in the Figure 4). When annealed at 300 ℃, the situation of all phases is also keep the same, but the peaks of ZrAl$_2$<200> and ZrAl$_2$<103> are much sharper. From 300 ℃, the interfacial structure is much different in two Al/Zr systems. After 1hr at 305 ℃ of Al(1%wtSi)/Zr (Figure 4(e)), the peak of Al<111> is smaller, and the ZrAl$_2$<200> is disappeared. Two new phases tetragonal <Zr,Al,Si><112> and tetragonal <Zr,Al,Si><103> appears around 38.79°. When the temperature increased to 500 ℃ in Al(1%wtSi)/Zr (Figure 4(g)), both tetragonal <Zr,Al,Si><103> and Zr<101> are disappeared, and the <Zr,Al,Si><112> and ZrAl$_2$<103> are much sharper. But the Zr<002> and Al<111> still have small peaks in the curve. For the curves in the Al(Pure)/Zr (Figure 4(f, i)), the peak of Al<111> becomes smaller, and the ZrAl$_2$<200> is also disappeared at 305 ℃. The new phase tetragonal ZrAl3<114> appears and shifts the peak position to 38.59°. At 500 ℃, the Zr<002> and Al<111> are still exist, but the ZrAl$_3$<114> and ZrAl$_2$<103> are much sharper, which means the new interfacial material could not destroy the multilayer structure completely. Based on the fitting data of Al(1%wtSi)/Zr, we found the interfacial phases transform from amorphous Al-Zr alloy to amorphous Al-Zr alloy and cubic ZrAl$_3$ after 290 ℃. When at 298 ℃, the interfacial phase is become to hcp ZrAl$_2$ and cubic ZrAl$_3$, which could smooth the interface (Figure 1(a)) and lower the surface roughness (Figure 3(a)). Up to 500 ℃, the multilayer structure still exists. Compared with Al(1%wtSi)/Zr and Al(Pure)/Zr, the Si in Al layer extends the reaction temperature from 295 to 298 ℃, and influences the composition of the interfacial material after 300 ℃. This phenomenon also happened in other systems, such as Ni/amorphous SiC multilayers [35]. Before 400 ℃, the interface is amorphous. After 400 ℃, the NiSi and Ni$_2$Si appeared at interface. Based on the AFM and XRD, the surface roughness increased with annealing temperature, and was correlated with similar increases in the interface width, which impact structural performance of the multilayers.

To verify the fitting results of XRD in the annealing process, TEM observations were carried out. Five specimens (RT, 295 ℃, 298 ℃, 300 ℃, 500 ℃ for Al(1%wtSi)/Zr systems) were made and performed on high magnification transmission electron micrographs (TEM) and selected area electron diffraction (SAED) pattern (The microstructure of Al(Pure)/Zr may be similar, which are not shown in Figure 5). In Figure 5(a), there is clearly an amorphous Al-Zr intermixed layer at each interface in the RT samples. The interlayers are symmetric, which the thicknesses are 1.5nm. The SAED pattern taken from the corresponding through-foil specimen (Figure 5(g)) shows the polycrystalline rings from cubic Al and an inner polycrystalline ring from hcp Zr. After 1h at 295 ℃, the asymmetric interlayers (Zr-on-Al and Al-on-Zr) are shown in the Figure 5(b), which the average thicknesses are 1.9nm and 1.7nm, respectively. When annealed at 298 ℃, the interlayers become clearly at interfaces, which smooth the interfacial boundary. The SAED pattern from the corresponding through-foil TEM sample (Figure 5(h)) show several faint new polycrystalline rings. The fringe spacing in the intermixed area are consistent with the ZrAl2<103> (d=0.2445nm), ZrAl2<200> (d=0.2278nm) and Al<111>/ZrAl3<111> (d=0.2320nm). After 1h at 300 ℃ (Figure 5(d)), the interdiffusion becomes larger than before 300 ℃. The thickness of Zr-on-Al interlayer (2.4nm) is larger than that of Al-on-Zr interlayer (2.0nm). By 500 ℃ in Figure 5(e, f), although the intermediate material can be found in the multilayers, the multilayer structure still exist, which the interlayers become much larger. The SAED in Figure 5(i) shows several polycrystalline rings, which the <Zr,Si,Al><112> (d=0.2348nm) and ZrAl2<103> can be easily observed in the pattern. Based on the analysis of TEM, the thicknesses of interlayers are consistent with the fitting data in the GIXR [13]. At 298 ℃, the interfacial boundary is very clearly, which decrease the crystal size (Figure 3(b)), and smooth the interfacial (Figure 1(a)) and surface roughness (Figure 3(a)). When annealed to 300 ℃, the asymmetric interlayers can be observed clearly, which the $ZrAl_2$ and $ZrAl_3$ can be found in the multilayers. Up to 500 ℃, the multilayer structure is not destroyed completely by the intermediate material. And the thickness of Zr-on-Al interlayer is much larger than that of Al-on-Zr interlayer. From the SAED, the fitting results of XRD can be verified, which we can find all possible phases in the corresponding patterns.

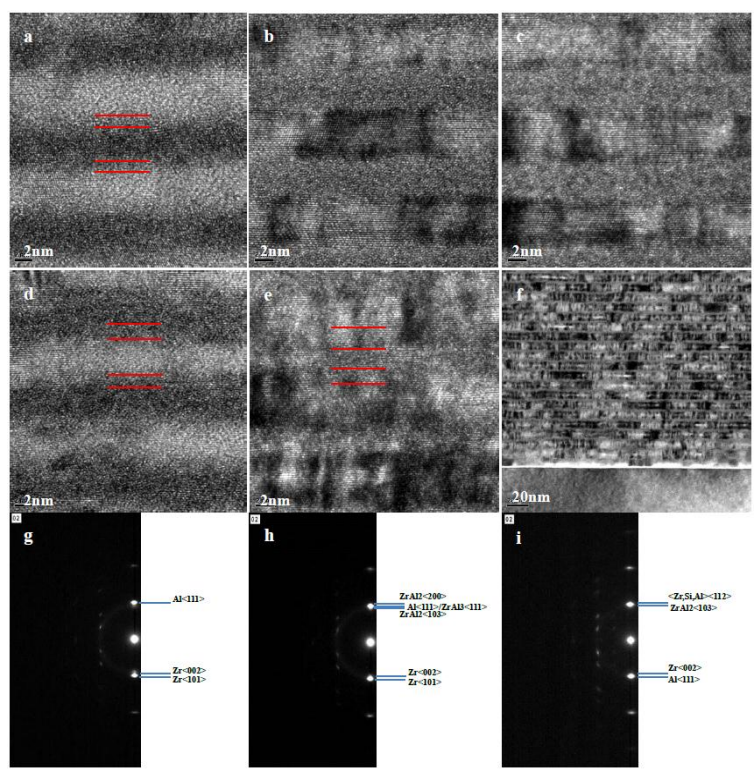

Figure 5. High magnification transmission electron micrographs and selected area electron diffraction pattern used to observed the cross-section of five Al(1%wtSi)/Zr multilayers (Al layer-lighter, Zr layer-darker). The micrographs and SAED patterns of the specimens RT (a, g), 295 ℃ (b), 298 ℃ (c, h), 300 ℃ (d), 500 ℃(e-2nm, f-20nm, i) are shown in the figure.

**3.3 Discussion**

To present transformation from symmetric to asymmetric in the annealing process for other systems , we describe the situation in Al/Zr systems firstly. At room temperature, the interlayers are symmetric because of the same binding energies for Al (fcc) and Zr (hcp) in the multilayers [36]. Although there is interdiffusion between Al and Zr, there is no new material appeared in the interface before 250 ℃, and the interdiffusion could not destroy the balance of the symmetry in the interlayers. After 250 ℃, too much amorphous Al-Zr alloy grows into Al and Zr layers, which induced the structure modification in the interface. That is Zr diffuses and reacts more rapidly with Al than Al diffuses and reacts with Zr [15]. The symmetric interlayers begin to transform to asymmetric interlayers in the multilayer. At higher temperature, the asymmetric become more obviously, which the new materials $ZrAl_2$ and $ZrAl_3$ appear in the multilayers, and enlarge the differences between two interlayers. Therefore, the explanation is suitable for the multilayers which have the symmetric interlayers at room temperature. And the interaction in the multilayers is also needed. This kind of multilayers might be Fe/Si multilayers and some alloy multilayers. For the Fe/Si multilayers, when the thickness of Fe is 4nm, Si-on-Fe interlayer is wider than Fe-on-Si interlayer, because the new compounds, FeSi and $FeSi_2$, appear in the interface [21]; While there is also shows that Fe-on-Si interlayer has more diffusion than Si-on-Fe interlayer at the thickness (2nm) of Fe, which Fe-on-Si interlayer could be $FeSi_2$ in the multilayer [36]. Thus, we can assume that there may have a symmetric interface in the Fe/Si multilayers, which the thickness of Fe could be in the range of 2-4nm. Using the annealing process to estimate the multilayers, we think there may have the similar situation with Al/Zr multilayers, which the symmetric interlayers could transform to asymmetric in the Fe/Si multilayers. For the alloy multilayers, they may also have the same symmetric performance with Al/Zr multilayers. Because the same binding energies for different layers could appear in the alloy multilayers, there may have symmetric interlayers. And the interdiffusion between two layers could also happen in the multilayers. The symmetric interlayers could change to asymmetric in the annealing process. Consequently, the symmetric performance in Al/Zr multilayers would also use to explain those kinds of structure modification during the annealing process.

## 4. CONCLUSION

We report the structural stability and interactions between Al and Zr layers in both Al(1%wtSi)/Zr and Al(Pure)/Zr multilayers. The thirteen multilayer samples for each system were deposited and annealed from 200 to 500 ℃ in a vacuum furnace for 1 h. To estimate the different reaction temperatures in the multilayers, the GXIR, AFM, XRD and TEM measurements are used. Based on the analysis, each system has three reaction temperatures before 300 ℃. At 250 ℃, the interlayer becomes asymmetric. The width of Zr-on-Al interlayer is larger than that of Al-on-Zr interlayer. Then the interfacial phases transform from amorphous Al-Zr alloy to amorphous Al-Zr alloy and cubic $ZrAl_3$ at 290 ℃. When annealed to 298 ℃ for Al(1%wtSi)/Zr and 295 ℃ for Al(Pure)/Zr, the interface become polycrystalline ($ZrAl_2$ and $ZrAl_3$) totally, which also have lower surface and interfacial roughness. The comparison of the two systems shows that the structural performance of Al(Pure)/Zr multilayers is much worse than that of Al(1%wtSi)/Zr in the whole annealing process. The Si in Al layer could not only delay the reaction temperature from 295 to 298 ℃, but also influence the composition of the interfacial material after 300 ℃. At 305 ℃, two new phase <Zr, Al, Si><112> and tetragonal <Zr, Al, Si><103> appear in Al(1%wtSi)/Zr, and only one phase $ZrAl_3$<114> appears in Al(Pure)/Zr multilayers. By 500 ℃, the multilayer structure still exists in both systems, which are consistent with the GIXR results and TEM images, and the asymmetric interlayers are more obviously in the multilayers. The transformation from symmetric to asymmetric in the annealing process for other systems is also present. Therefore, based on the analysis, the Al/Zr multilayers have a stable structure and performance up to 290 ℃, which have no chemical reactions at interfaces.

## ■ REFENRENCE: